# Optical spectra of quantum dots: effects of non-adiabaticity

**J. T. Devreese**[*,1], **V. M. Fomin**[1,2], **E. P. Pokatilov**[2], **V. N. Gladilin**[1,2], and **S. N. Klimin**[1,2]

[1] Theoretische Fysica van de Vaste Stoffen, Universiteit Antwerpen (UIA), Universiteitsplein 1,B-2610 Antwerpen, Belgium

[2] Fizica Structurilor Multistratificate, Universitatea de Stat din Moldova, str. A. Mateevici 60, MD-2009 Chisinau, Moldova

It is shown that in many cases an adequate description of optical spectra of semiconductor quantum dots requires a treatment beyond the commonly used adiabatic approximation. We have developed a theory of phonon-assisted optical transitions in semiconductor quantum dots, which takes into account non-adiabaticity of the exciton-phonon system. Effects of non-adiabaticity lead to a mixing of different exciton and phonon states that provides a key to the understanding of surprisingly high intensities of phonon satellites observed in photoluminescence spectra of quantum dots. A breakdown of the adiabatic approximation gives an explanation also for discrepancies between the serial law, observed in multi-phonon optical spectra of some quantum dots, and the Franck-Condon progression, prescribed by the adiabatic approach.

**1 Introduction** Quantum dots continue to be an area of intense scientific activity, because they have a number of advantages as the building stones for advanced semiconductor devices with three-dimensional band-structure engineering. Considerable effort is being devoted to the investigation of effects due to the exciton-phonon interaction on optical properties of quantum dots. Numerous photoluminescence and Raman measurements on quantum dots (for instance, on CdSe and PbS spherical nanocrystals [1-3] as well as on InAs/GaAs self-assembled quantum dots [4-8]) reveal surprisingly high probabilities of phonon-assisted optical transitions. In InAs/GaAs self-assembled quantum dots, the so-called experimental Huang-Rhys factor, determined as $S_K^{(\exp)} \equiv K I_K / I_{K-1}$ with $I_K$, the measured intensity of the $K$-phonon satellite, reaches values 0.05 to 0.5 [4-7]. The values of $S_K^{(\exp)}$, measured by different groups, show a well-pronounced trend to an increase with diminishing size of quantum dots [6].

Attempts to interpret some of the above experiments on the basis of the adiabatic theory [9,10] meet difficulties. Thus, for spherical CdSe quantum dots, the calculated Huang-Rhys factor $S$ takes values, which are by one or two orders of magnitude smaller than the experimental Huang-Rhys factor $S_K^{(\exp)}$. Moreover, some experimental spectra cannot be described by a Franck–Condon progression, predicted by the adiabatic theory, even when the Huang-Rhys factor is considered as a fitting parameter. In the framework of the adiabatic approach, various mechanisms, which ensure separation of the electron and hole charges in space, are commonly considered as a possible origin for the increased Huang-Rhys factor [4,11]. We show that the non-adiabatic treatment of phonon-assisted optical transitions, which was proposed in Ref. [12], can provide an explanation for high intensities of phonon satellites observed in optical spectra of semiconductor quantum dots as well as for unusual serial laws, obeyed by these intensities.

---
[*] Corresponding author: devreese@uia.ua.ac.be, Phone: +32 3 820 2459, Fax: +32 3 820 2245

**2 Light absorption by quantum dots** Following the approach used in Ref. [12] and based on the Kubo formula for the optical response, the linear coefficient of absorption by the exciton-phonon system in a quantum dot is described by the expression

$$\alpha(\Omega) \propto \mathrm{Re} \sum_{\beta,\beta'} d_\beta^* d_{\beta'} \int_0^\infty dt\, e^{i(\Omega-\Omega_\beta+i\varepsilon)t} \langle \beta | \overline{U}(t) | \beta' \rangle, \tag{1}$$

where $\Omega$ is the light frequency, $d_\beta$ and $\Omega_\beta$ are, respectively, the dipole matrix element and the Franck-Condon frequency of a transition between the exciton vacuum state and the one-exciton state $|\beta\rangle$. The evolution operator averaged over the phonon ensemble, $\overline{U}(t)$, is represented in terms of the chronological ordering operator T:

$$\overline{U}(t) = \mathrm{T}\exp\left\{-\frac{1}{\hbar^2}\sum_\lambda \int_0^t dt_1 \int_0^{t_1} dt_2 \left[(\overline{n}_\lambda+1) e^{-i\omega_\lambda(t_1-t_2)}\gamma_\lambda(t_1)\gamma_\lambda^+(t_2) + \overline{n}_\lambda\, e^{i\omega_\lambda(t_1-t_2)}\gamma_\lambda^+(t_1)\gamma_\lambda(t_2)\right]\right\}. \tag{2}$$

In Eq. (2), the index $\lambda$ labels phonon modes, $\omega_\lambda$ are phonon frequencies, $\gamma_\lambda(t)$ are the exciton-phonon interaction amplitudes in the interaction representation, and $\overline{n}_\lambda = [\exp(\hbar\omega_\lambda/k_\mathrm{B}T)-1]^{-1}$.

Within the adiabatic approximation, which is widely used to treat optical spectra of quantum dots, non-diagonal matrix elements of the exciton-phonon interaction are neglected when calculating the absorption coefficient given by Eq. (1) with Eq. (2). The adiabatic approach [9,10] supposes that (i) both the initial and final states for a quantum transition are non-degenerate, (ii) energy differences between the exciton states are much larger than the phonon energies. We have revealed that these conditions are often violated for optical transitions in quantum dots. In other words, the exciton-phonon system in a quantum dot can be essentially *non-adiabatic*. In Ref. [12], we have developed an approach to calculate the absorption coefficient given by Eqs. (1) and (2) taking into consideration non-adiabaticity.

Figure 1 illustrates the crucial role of non-adiabaticity in the phonon-assisted light absorption by InAs/GaAs quantum dots. The absorption spectrum is calculated using a Green-function approach, which allows us to take into account the non-adiabaticity effects on both the transition probabilities and the energy spectrum of the exciton-phonon system. We have modelled the quantum-dot structure by an InAs cylinder of height $h$ and radius $R$ embedded into GaAs. The absorption spectrum is calculated taking into account four lowest energy levels of an electron-hole pair. Only two of these levels are optically active in the dipole approximation. Within the adiabatic approximation, only the peaks corresponding to the zero-phonon transitions to the optically active states are clearly seen, while the intensities of the phonon satellites appear to be negligibly small. The electron-phonon interaction mixes different optically active and non-active states. This non-adiabatic mixing results in a rather rich structure of the absorption spectrum of the exciton-phonon system [13,14]. Similar conclusions about a dramatic influence of the exciton-phonon interaction on the optical spectra of quantum dots have been recently formulated in Ref. [15] in terms of a strong coupling regime for excitons and LO phonons. Such a strong coupling regime is a particular case of the non-adiabatic mixing related to a (quasi-) resonance when the spacing between exciton levels is close to the LO phonon energy. The above results imply that, due to the effects of non-adiabaticity, an enhancement of phonon-satellite intensities must occur also in resonant-photoluminescence and photoluminescence excitation (PLE) spectra of quantum dots. New experimental evidence of the enhanced phonon-assisted absorption due to effects of non-adiabaticity has been recently provided by PLE measurements on single self-assembled InAs/GaAs [8] and InGaAs/GaAs [16] quantum dots.

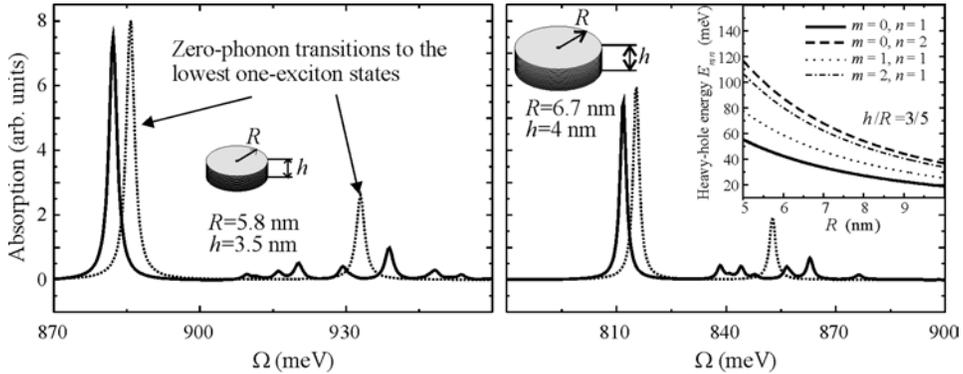

**Fig. 1** The absorption spectra calculated within the adiabatic approximation and the non-adiabatic approach for cylindrical InAs quantum dots with different sizes. Inset: four lowest energy levels for heavy holes. $m$ is the projection of the angular momentum on the cylinder axis, $n$ is related to the quantization of the radial motion.

**3 Photoluminescence and Raman spectra of quantum dots** The photoluminescence spectra are strongly influenced by relaxation processes during the time interval between excitation and recombination of an exciton. In Fig. 2, we show the thermodynamic equilibrium photoluminescence spectra of spherical CdSe quantum dots at various temperatures. Over a wide range of radii of these quantum dots, the magnitude of the splitting of the lowest exciton level by the crystal field is close to the LO phonon energy. The non-radiative energy relaxation between the exciton energy sublevels, in combination with effects of non-adiabaticity, lead to a substantial difference of the multiphonon photoluminescence spectra from the Franck-Condon progression. Indeed, for the measured $K$-phonon-peak intensities, $I_K$, the ratios $S_1^{(\exp)}$ and $S_2^{(\exp)}$ obviously differ from each other. As seen from Fig. 2, taking non-adiabaticity into consideration allows for an adequate description of the observed phonon-peak intensities as well as of their dependence on temperature. Also the serial laws obeyed by the measured intensities of phonon satellites, find a satisfactory interpretation within our approach.

A theory of the resonant Raman scattering in semiconductor quantum dots is developed taking into account the non-adiabaticity of the exciton-phonon system. Using the Hamiltonian of the electron-phonon interaction, derived within the multimode dielectric continuum model [17], we treat resonant Raman scattering via the multiphonon exciton transitions. The aforesaid model of phonons exploits both electrostatic and mechanical boundary conditions for the relative ionic displacement vector, as well as the phonon spatial dispersion in bulk. The confined phonon modes in a quantum dot are hybrids of bulk-like and interface phonons. With non-adiabaticity taken into account, Raman spectra calculated for CdSe and PbS spherical quantum dots [17] agree well with the experimental data [1,3].

We have applied our theoretical technique to GaAs/AlAs quantum disks with typical geometrical parameters from Ref. [18]. The one-phonon and two-phonon Raman spectra are shown in Fig. 3. The "GaAs-band" calculated within the present model reveals a structure similar to that of the experimental Raman spectra of quantum disks [18] (in the inset). Effects of non-adiabaticity considerably enhance both the absolute values of Raman peak intensities and the relative intensities of the two-phonon peaks with respect to those of the one-phonon peaks.

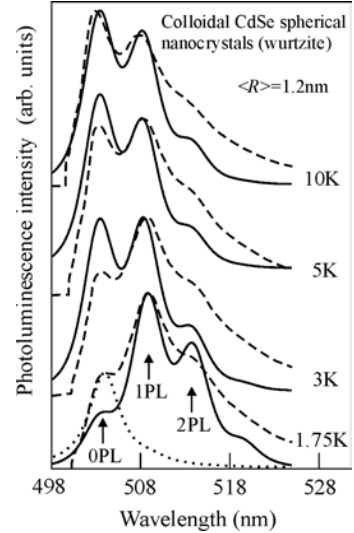

**Fig. 2** Photoluminescence spectra of spherical CdSe quantum dots with the average radius $\langle R \rangle$ =1.2 nm at various temperatures. The dashed curves represent the experimental data [2], the dotted curve displays a Franck-Condon progression with the Huang-Rhys factor $S = 0.06$ calculated in Ref. [11] and the solid curve results from our theory. $N$-phonon lines ($N$PL) with $N$ = 0, 1, 2 are indicated by arrows.

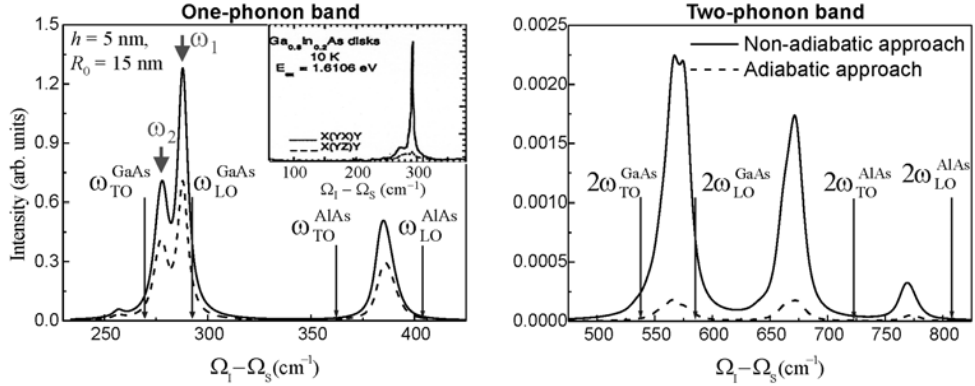

**Fig. 3** One-phonon and two-phonon Raman spectra for a cylindrical GaAs/AlAs quantum dot with the height $h$ = 5 nm and with the lateral-confinement frequency parameter $\Omega_0$ = 3 meV. Solid curves are obtained taking into account effects of non-adiabaticity. Dashed curves correspond to the adiabatic approximation. Inset: measured Raman spectrum of a disk-shape self-assembled $Ga_{0.8}In_{0.2}As$ quantum dot (from Ref. [18]).

**4 Conclusions** We have shown that non-adiabaticity is an inherent property of exciton-phonon systems in various quantum-dot structures, which drastically enhances the efficiency of the exciton-phonon interaction. Effects of non-adiabaticity are of paramount importance when treating remarkably high intensities of phonon peaks observed in optical spectra of quantum dots. It is also shown that the non-adiabatic mixing of different exciton and phonon states leads to a rather rich structure of optical spectra as compared to those given by the adiabatic approximation.

**Acknowledgements** This work has been supported by GOA BOF UA 2000, IUAP, FWO-V project No. G.0274.01N, WOG WO.025.99N (Belgium), and the European Commission GROWTH Programme, NANOMAT project, contract No. G5RD-CT-2001-00545.